# AirTag, You're It: Reverse Logistics and Last Mile Dynamics

**David A. Noever and Forrest G. McKee**
**PeopleTec, Inc., Huntsville, AL**
david.noever@peopletec.com        forrest.mckee@peopletec.com

**ABSTRACT**

This study addresses challenges in reverse logistics, a frequently overlooked but essential component of last-mile delivery, particularly in disaster relief scenarios where infrastructure disruptions demand adaptive solutions. While hub-and-spoke logistics networks excel at long-distance scalability, they often fail to optimize closely spaced spokes reliant on distant hubs, introducing inefficiencies in transit times and resource allocation. Using 20 Apple AirTags embedded in packages, this research provides empirical insights into logistical flows, capturing granular spatial and temporal data through Bluetooth LE (BLE) 5 trackers integrated with the Apple Find My network. These trackers demonstrated their value in monitoring dynamic cargo movements, enabling real-time adjustments in mobile hub placement and route optimization, particularly in disaster relief contexts like Hurricane Helene. A novel application of discrete event simulation (DES) further explored the saddle point in hub-spoke configurations, where excessive hub reliance clashes with diminishing spoke interaction demand. By coupling simulation results with empirical AirTag tracking, the study highlights the potential of BLE technology to refine reverse logistics, reduce delays, and improve operational flexibility in both routine and crisis-driven delivery networks.

**INTRODUCTION**

Bluetooth Low Energy (BLE-5) trackers, particularly Apple AirTags, have emerged as research tools for item tracking, sparking investigations into their functionality, privacy implications, and potential misuse. The present work surveys BLE-5 tags as practical tools for tracking mailed item. These devices rely on dense networks of smartphones for location updates, offering a decentralized passive sensor and energy-efficient tracking solution that contrasts sharply with traditional technologies like GPS or inertial measurement units (IMUs).  BLE-5 is a wireless communication protocol designed for low-power, short-range data transmission. It represents a significant advancement over earlier Bluetooth standards, particularly in terms of energy efficiency, data transmission rates, and range. BLE 5 supports a range of up to 240 meters in ideal conditions and offers data rates of up to 2 Mbps. Figure 1 summarizes the NFC pairing antenna (used to initially pair the AirTag to an Apple network device), the BLE-5 (longer range) and UWB (shorter-range) components for transmission. These tracking antennae make it highly suitable for Internet of Things (IoT) applications, where devices like AirTags need to operate over extended periods with minimal energy consumption.

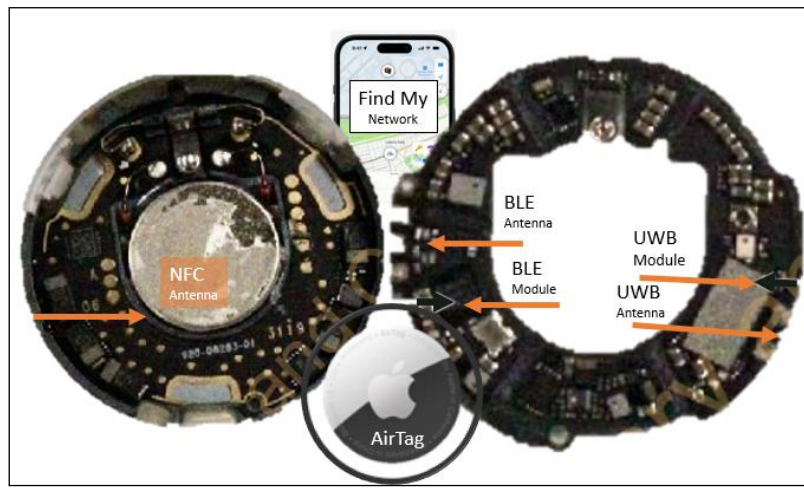

*Figure 1. Apple AirTag with 3 Antenna (NFC, BLE, UWB)*

Studies like Shafqat et al. [1] and Alamleh et al. [2] have outlined security vulnerabilities and defensive mechanisms, highlighting the balance between utility and risk [3-20]. Roth et al. [3] explored exploits involving modified trackers, while Ibrahim et al. [4] examined the broader adoption dynamics of BLE tracking systems. Further, McBride [5] and Hernández-Orallo et al. [6] evaluated AirTags as people-tracking devices and their efficiency in various scenarios. Recent works, such as those by Heinrich et al. [21], have emphasized anti-stalking measures and democratized protections for users, while others, like Gallagher [9], investigated innovative applications such as animal tracking. Owners have used AirTags to prevent car theft by hiding them in vehicles, while car thieves have exploited them to track and steal high-value targets. Fleet managers and neighborhood watches have used AirTags in reverse geofencing setups to monitor unfamiliar or suspicious delivery vehicles. In personal safety contexts, AirTags have been both misused by stalkers and employed as tools to detect and deter stalking through alarms and geofences. Airline passengers famously used AirTags to locate lost luggage more effectively than airlines, leading some carriers to adopt them as tracking aids or ban them. Investigators, both civilian and professional, have deployed AirTags to track recyclables and expose fraudulent waste claims, while activists used them to reveal a covert German intelligence agency operating under a proxy telecom address. In another instance, AirTags inadvertently revealed ground vehicle customs points for NATO supplies entering Ukraine, a detail later obscured in public media. These use cases underscore the transformative potential of AirTags beyond their intended purpose while also illuminating critical challenges around privacy, security, and ethical use.

This unique operational paradigm makes BLE trackers particularly intriguing when compared to more specialized tracking systems employed in scientific research, such as meteorological GPS-equipped tornado probes or wildlife tagging devices as detailed in Table 1. Each of these technologies is designed to reveal dynamic patterns in their respective domains, yet BLE trackers may offer advantages in urban logistics and mailing studies. Meteorological research has long relied on GPS-enabled probes to study the chaotic and high-speed dynamics of tornadoes, leveraging the precision and global coverage of satellite-based systems. However, these probes often encounter limitations in urban environments, where signal obstructions from buildings and infrastructure create blind spots. Similarly, in urban planning studies, tracking recyclable materials often requires GPS or radio-frequency identification (RFID) tags to trace waste streams through city systems. While effective in open environments, these methods can be costly and are vulnerable to interference in densely populated areas. BLE trackers, in contrast, capitalize on the proximity of ubiquitous smartphone networks, providing an inexpensive, scalable, and interference-resistant alternative for tracking packages or objects as they traverse complex logistical pathways. The market penetration for Apple iPhones reaches approximately 60% of all passerby opportunities or casual relays that might transmit a location to the central aggregator in this study.

By embedding AirTags in packages sent to and from remote or urban areas, researchers have previously explored the dynamics of delivery networks under real-world conditions [22-33], identifying inefficiencies [32-33] and testing the resilience of logistical pathways for international routes and within the US (44% of the world's mail or 116 billion deliveries in 2023 [33]). As an example of smart logistics, Korea has experimented with RFID stamps to enable fine-grained mail tracking outside of the standard inventory at hub processors. This approach parallels studies in animal movement and environmental systems but adapts the technology to the human-engineered systems that sustain modern commerce and communication. Through this lens, BLE trackers represent not only a practical tool for mailing studies but also a broader innovation in understanding and optimizing the flow of goods and resources in increasingly complex and interconnected environments. One outcome of the present work is to overcome the AirTag limitations to report only current locations without logging history and thus enable route optimizations and analyses.

| Delivery Factor | Patterns to Explore |
|---|---|
| **Dwell Time** | • Hub queuing duration vs. time of day |
| | • Loading/unloading patterns at spokes |
| | • Nub distance and correlated wait times |
| **Route Efficiency** | • Detour ratio: actual/direct distance |
| | • Backtracking frequency between spokes |
| | • Empty vehicle miles percentage |
| **Temporal Patterns** | • Peak vs. off-peak travel times |
| | • Seasonal variation in route selection |
| | • Time-of-day impact on hub accessibility |
| **Speed Profiles** | • Hub approach/departure velocities |
| | • Inter-spoke travel speeds |
| | • Congestion-related slowdowns |
| **Service Territory** | • Spoke service area overlap |
| | • Customer density distribution |
| | • Hub coverage zone efficiency |
| **Vehicle Utilization** | • Load factor vs. distance from hub |
| | • Capacity usage patterns |
| | • Multiple hub visit frequency |
| **Bottleneck Analysis** | • Recurring delay locations |
| | • Hub entrance congestion patterns |
| | • Spoke access constraints |
| **Environmental Impact** | • Weather-related route changes |
| | • Seasonal pattern adjustments |
| | • Alternative route frequency |

**Table 1. Variable Search Space of Reverse Logistics**

**METHOD**

The study used a set of 15 AirTags, each paired initially with an iPhone running iOS 18.1.1, to evaluate the capabilities of BLE 5 trackers in logistical scenarios. The AirTags [34] were monitored and logged automatically using Apple's Find My application on a MacBook Pro (2011) running macOS Monterey, leveraging its ability to generate unencrypted JSON logs for analysis. These logs, contained within the cached items.data file, provided rich metadata about each tracker, including location details, device attributes, and battery status (Table 1). The main challenges here involved acquiring the right macOS version, then automating the posting of every location change to a Google Drive folder while converting the raw device files to an aggregated Keyhole Markup Language (KML) layer for geographic display on 3D globes.

To distinguish between the devices, all tags took on the name of an historical or fictional explorer (Figure 2). The Find My app served as the primary interface for monitoring the trackers in real time, displaying their positional data and metadata. The unencrypted items.data file produced by the app captured key parameters such as geographic coordinates (latitude, longitude, altitude), floor level, and the accuracy of these measurements, along with temporal data like timestamps for each location update. As shown in Figure 2, after nearly 20,000 location updates, the mean frequency per device is 6.5 per hour and 117 for the network of 19 tracked items or devices studied. As shown in Figures 2-3, the distribution varies considerably by device name, type, and day depending on the surrounding UWB relay environment. For instance, the update frequency will vary by background traffic particularly for moving devices and by overall location. The stationary MacBook registers the highest update frequency as one might expect, followed by the registration iPhone (moving) and the disseminated AirTag trackers.

| Location Frequency | Overall | Per Tracker |
|---|---|---|
| Total Updates (7 day) | 19629 | 1091 |
| Updates per Day | 2804 | 156 |
| Updates per Hour | 117 | 6.5 |
| Updates per Minute | 1.95 | 0.11 |

Table 2. Location update frequency based on 18 tracked items and a 7-day observation window

Additional device-specific fields, including battery level, operational status, and model information, were also logged, offering comprehensive insights into the trackers' operational status during deployment. AirTags continuously broadcast encrypted identifiers (payloads) via BLE. These payloads include no personally identifiable information but serve as unique identifiers for the device. When a compatible Apple device—such as an iPhone, iPad, or Mac—comes into proximity of an AirTag, it captures the BLE signal and relays the encrypted identifier along with its own GPS coordinates to Apple's Find My servers. This process is entirely anonymous and end-to-end encrypted, ensuring privacy for the AirTag owner and the relay device's user.

The spatial data collected allowed the study to evaluate the AirTags' performance in tracking and logistical workflows, particularly their ability to provide accurate and timely updates in a simulated mailing network. By parsing the unencrypted JSON data, the study not only validated the reliability of BLE 5 trackers in dynamic environments but also uncovered their limitations, such as potential discrepancies in altitude or positional accuracy when navigating complex spatial environments like multistory buildings. The combination of real-time monitoring through the Find My app and retrospective analysis of the items.data file created an empirical framework for understanding the dynamics of BLE-based tracking in logistical systems.

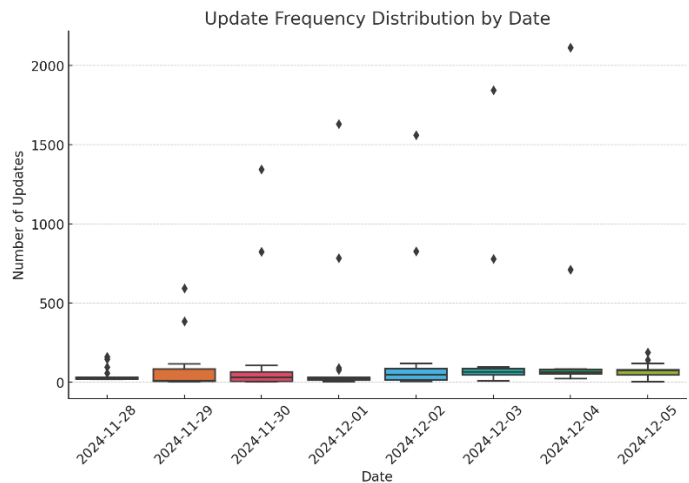

*Figure 2. Weekly observation variability by day and outliers for tracked items as outliers*

**RESULTS**

*Last Mile Problem.* In spoke and hub models, the logistics of sending items to closely spaced spoke mailboxes add considerable travel, fuel, and time delay when all sorting happens at a distant hub. This network models shares features with traditional the last-mile problem. The last mile, as defined in supply chain management, refers to the final segment of delivery from a transport hub to the recipient's destination. This stage is often the most complex and resource-intensive part of the delivery process, especially in disrupted or underserved areas where infrastructure may be damaged or inadequate. BLE trackers have the potential to serve as "last mile probes," providing real-time insights into the dynamics of this critical stage and helping optimize service delivery in challenging environments.

Reverse logistics, an often-overlooked aspect of last-mile delivery, was also explored by tracking the return-to-sender processes for undeliverable packages. AirTags embedded in these packages (Figure 3) provided visibility into the efficiency and timeliness of returns, offering insights into how mobile postal units can better manage reverse logistics during disaster recovery operations. The integration of AirTag data with geographic visualization tools, such as Google Maps, allowed for the creation of detailed route maps, highlighting inefficiencies and bottlenecks in the delivery network.

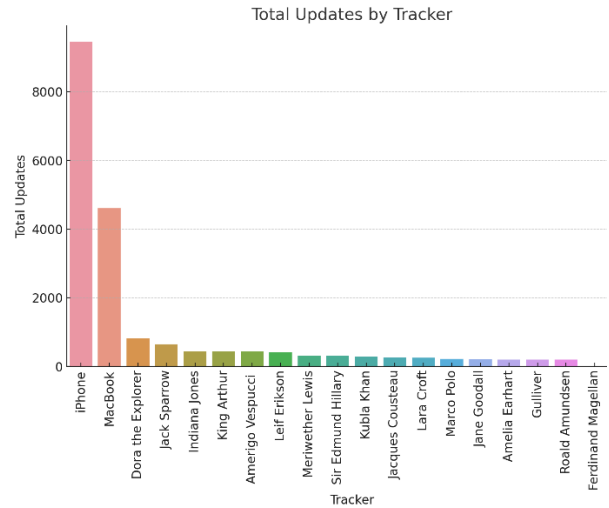

*Figure 3. Frequency distribution across devices (MacBook, iPhone) vs. relay items (named AirTags) over a week*

*Theoretical Motivation for Hub-Spoke Network Analysis.* The optimization of hub-and-spoke networks presents a complex interplay between spatial configuration and operational dynamics. The system's total cost function captures three critical components:

$$C(d_s, d_h, \lambda) = \alpha \cdot C_{travel}(d_s, d_h, \lambda) + \beta \cdot C_{time}(d_s, d_h, \lambda) + \gamma \cdot C_{congestion}(d_h, \lambda)$$

where:
- $C_{travel}$ represents distance-based costs proportional to $(2d_h + d_s)$
- $C_{time}$ captures delays from both travel and processing
- $C_{congestion}$ models non-linear queuing effects as $\lambda$ approaches hub capacity $\mu$

Discrete Event Simulation (DES) is particularly well-suited for analyzing this system due to several key characteristics [35]. First, the queuing dynamics at the hub exhibit complex behavior that resists closed-form analytical solutions, especially when arrival rates approach capacity limits. Second, the interaction between spatial distance and demand follows an inverse square relationship ($\lambda \propto 1/d_s^2$), creating non-linear effects that are best understood through simulation. Finally, the system's response to perturbations in arrival rates and processing times creates temporal dynamics that are difficult to capture in static optimization models.

The DES approach enables exploration of critical pinch points in the network where operational constraints meet service demands. By modeling mail generation as a Poisson process and implementing realistic queue management at the hub, the simulation reveals how congestion effects propagate through the system. This understanding is crucial for identifying optimal hub-spoke distances that balance centralization benefits against increasing travel costs and diminishing interaction demand over distance.

The key insight from this theoretical framework is that hub-spoke networks exhibit a saddle point in their cost function: at small spoke distances $d_s$, the hub becomes inefficient due to excessive detours, while at large distances, interaction demand $\lambda$ drops too low to justify centralization. Finding this optimal configuration through simulation provides practical guidance for logistics network design while accounting for the stochastic nature of real-world operations.

The discrete event simulation highlights critical inefficiencies in traditional hub-and-spoke logistics systems, particularly when two spokes are closely spaced (Figure 4). In the simulated scenario, all mail between two nearby spokes, Alpha and Beta, was routed through a centralized hub located 50 miles away. This hub dependency led to significant delays as mail had to travel an unnecessary detour of nearly 100 miles, despite the spokes being only 5 miles apart. The inefficiency of this routing system becomes apparent when considering the compounded travel times and processing delays that occur at the hub.

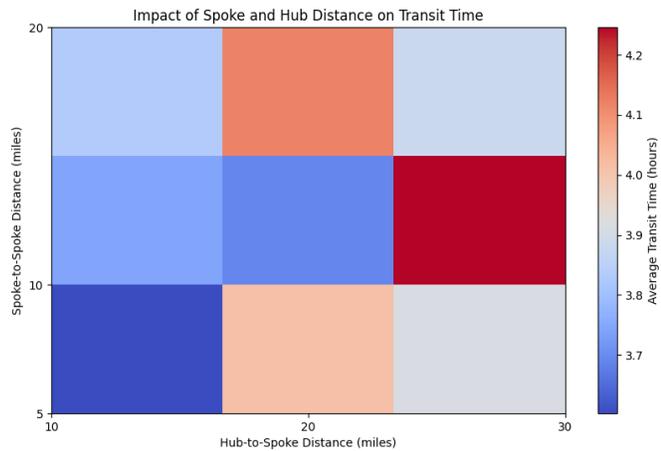

*Figure 4. Discrete Event Simulation Results for Parameter Space Exploration for Two Close Spokes and a Distant Hub*

The simulation revealed that closely spaced spokes exacerbate the inherent limitations of hub-and-spoke systems. While the central hub is effective for aggregating and redistributing mail over long distances, it is a poor fit for localized delivery needs between spokes in proximity. The forced reliance on the hub introduced avoidable delays, as vehicles prioritized the hub's queue rather than addressing the short and direct routes between Alpha and Beta. This bottleneck effect became more pronounced during peak times, when incoming mail at the hub exceeded the processing and vehicle capacity.

Temporal patterns in the simulation illustrated additional inefficiencies, as mail often arrived at the hub faster than it could be dispatched to its destination. This queuing problem delayed deliveries further, with mail from Alpha to Beta spending unnecessary time waiting for available transport, despite the short geographic distance between the two. For time-sensitive deliveries, such delays would be particularly costly, as they could result in missed deadlines or reduced customer satisfaction.

The findings suggest that a centralized hub approach is fundamentally misaligned with the dynamics of localized delivery between nearby spokes. The model demonstrates the potential benefits of introducing direct spoke-to-spoke routing, which could significantly reduce travel distances and delivery times. Such routing would bypass the hub entirely for Alpha-Beta exchanges, allowing for more efficient mail flow and less dependency on the hub's processing capabilities. Additionally, decentralizing operations by establishing mini-hubs closer to spoke clusters could mitigate the detour effect and reduce overall delays.

By illustrating the inefficiencies of hub-dominated logistics in last-mile delivery scenarios, the simulation in Figure 5 emphasizes the importance of adaptive routing strategies. Combining centralized and decentralized methods based on the geographic relationship between spokes and the hub could optimize system performance, particularly for high-frequency routes between closely spaced locations. The results demonstrate that while hub-and-spoke systems offer scalability for long-distance logistics, their rigidity limits their effectiveness in localized, dynamic delivery networks. This insight is crucial for designing more flexible systems that account for the specific spatial and temporal challenges of the last mile.

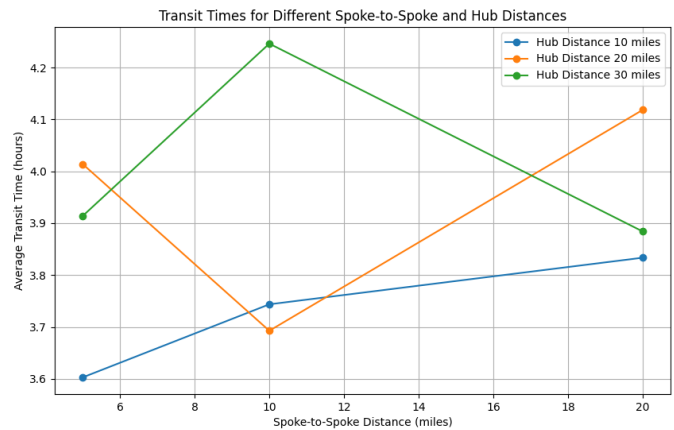

*Figure 5. Complex Transit Time Dynamics with Spoke-Spoke Distance vs. Hub-Spoke Distance*

*Hurricane Helene and western North Carolina relief (2024).* In the context of Hurricane Helene, where western North Carolina faced extensive road blockages and community displacement, AirTags could provide granular data on the movement of relief supplies through temporary postal services. For example (Figure 6), mobile units stationed at pop-up distribution hubs or traveling between evacuation centers could be equipped with BLE trackers to monitor their location and efficiency in real-time. This data could then be fed into centralized logistics systems, enabling disaster response teams to dynamically allocate resources, reroute supplies around inaccessible areas, and ensure that critical deliveries reach high-need locations. Like real-time trackers from the state emergency management system, the postal tags provide route information and identify bottlenecks and pinch points. An overlay of road closures and power outages support the logistics network with road CCTV and updated maps of flooded areas, shelters, evacuation routes, fuel, and emergency staging areas.  The study hypothesized that AirTags would provide consistent and accurate location data, even in challenging environments where traditional GPS systems often fail. Their reliance on the dense Find My network, which utilizes proximity to smartphones rather than satellite signals, was expected to ensure reliable updates in urban canyons, dense forests, or temporary shelters where GPS may be obstructed. Furthermore, the data was analyzed to identify delays and deviations in delivery routes caused by infrastructure disruptions. For instance, packages encountering detours or delays at flooded roads were tracked in real time, allowing for dynamic rerouting decisions and resource reallocation (Figure 7).

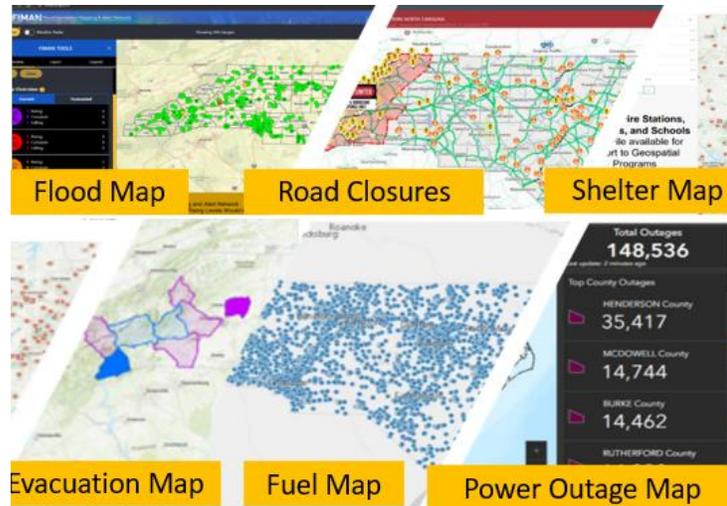

*Figure 6. Typical EMS Requested Disaster Needs for Real-Time Tracking*

**DISCUSSION**

The Find My network offers distinct advantages for reverse logistics compared to traditional tracking methods, particularly in scenarios where efficiency, cost, and adaptability are critical. Unlike dedicated GPS-based systems, which require consistent satellite connectivity and often involve high operational costs, Find My relies on the vast network of Apple devices to relay location data. This crowdsourced architecture enables the tracking of items without the need for direct GPS integration, making AirTags a lightweight and cost-effective alternative for monitoring reverse logistics.

In reverse logistics scenarios, such as return-to-sender processes or undeliverable packages, AirTags provide robust tracking capabilities by leveraging Bluetooth Low Energy (BLE) technology to send location updates through nearby Apple devices. This approach is particularly effective in urban and suburban environments, where device density ensures frequent updates. By contrast, GPS-based trackers may outperform AirTags in sparsely populated or rural areas with limited Apple device penetration, as the Find My network's reliance on proximity introduces potential gaps in coverage. However, GPS systems are more power-intensive and less practical for widespread deployment in cost-sensitive logistics operations.

The results indicate that while AirTags excel in tracking the movement of items through complex urban routing, their performance is influenced by the density of the surrounding Apple ecosystem. Reverse logistics operations benefit from AirTags' ability to identify bottlenecks and delays at specific distribution hubs, as shown by consistent location updates in high-density areas. However, in low-density zones, the lack of proximate Apple devices can result in delayed updates, highlighting a tradeoff between affordability and real-time precision.

In terms of flexibility, the Find My network allows for seamless integration into existing workflows without the need for proprietary infrastructure, unlike many GPS-based systems that require specialized hardware or software. While both systems have unique strengths, AirTags and the Find My network stand out as a scalable, energy-efficient option for reverse logistics in environments where device density and cost constraints are prioritized over precision and continuous updates. It is worth noting that significant portions of the world's population reside in regions with varying degrees of GPS and satellite technology restrictions, encompassing approximately 22% of global landmass and 43% of the world's population. This includes major nations such as China, India, Pakistan, Iran, Russia, and others, where GPS technology faces regulatory hurdles ranging from registration requirements to outright bans. In this context, Bluetooth-based tracking technologies like AirTags present an alternative tracking solution that operates independently of GPS infrastructure. These devices leverage crowd-sourced networks of consumer devices to provide location data, potentially circumventing satellite-based restrictions while still enabling basic location tracking capabilities.

A significant limitation encountered during the research was the integration of encryption in Apple's Find My network, which affects the accessibility of raw data from AirTags. AirTags leverage advanced encryption to ensure user privacy and data security, making it difficult to directly access their internal location metadata or transmission details. However, this limitation was circumvented by employing a legacy MacBook Pro (2011 model) running macOS Monterey, a system version that allowed access to unencrypted logs from the Find My application. macOS Monterey, released on October 25, 2021, introduced substantial features and compatibility enhancements for Macs produced in 2015 or later, yet its support for legacy systems was invaluable for this study. The MacBook provided an environment where the items.data JSON logs could be accessed without encryption, enabling the analysis of AirTag metadata, including location, battery status, and positional accuracy. This workaround highlighted the potential barriers researchers face when modern security protocols obscure device telemetry, posing challenges for detailed experimental setups.

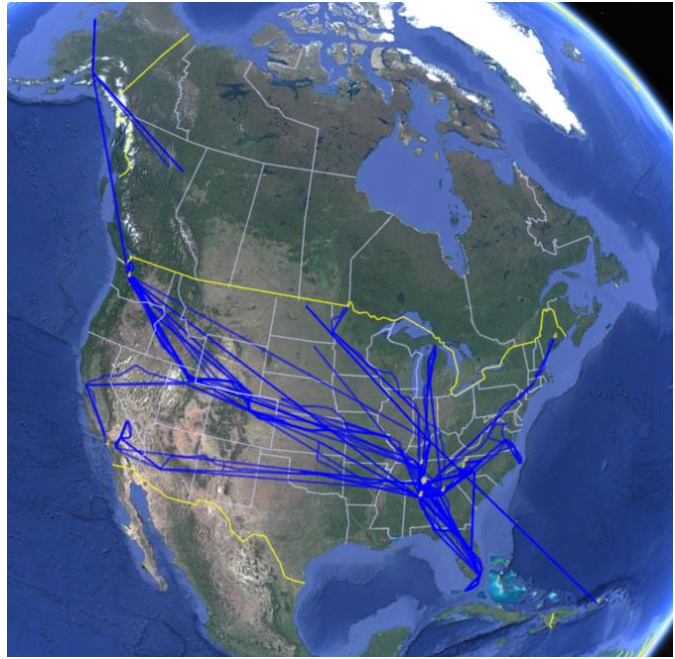

*Figure 7. Mail traces similar to network pings through hub and spoke delivery networks across the US Postal Service*

In contrast to Apple's AirTags, Samsung SmartTags operate on a more traditional Bluetooth framework and lack a similarly extensive network backbone to pivot signals when the primary user is out of range. SmartTags rely on direct connections to the owner's device or other Samsung Galaxy devices within the SmartThings ecosystem. While effective in close-range scenarios, this architecture significantly limits their utility in large-scale or geographically distributed tracking operations. The absence of a global crowdsourced network like Apple's "Find My" reduces the range and frequency of updates, making SmartTags less suitable for scenarios requiring continuous location monitoring or dynamic routing adjustments. For instance, if a SmartTag moves out of range of nearby compatible devices, its location cannot be updated until it re-enters a networked area, creating potential blind spots in tracking.

The comparison underscores the unique advantages of Apple's Find My network, which mitigates such gaps by using the extensive base of Apple devices as signal relays. While SmartTags offer a simpler, more direct Bluetooth tracking system, their reliance on the narrower Samsung ecosystem restricts their effectiveness for applications like reverse logistics or disaster relief, where wide coverage and decentralized signal propagation are critical. This difference highlights the importance of network architecture in determining the scalability and reliability of BLE-based tracking systems.

**CONCLUSION**

This research explored the dynamics of last-mile delivery in hub-and-spoke logistics systems using BLE trackers such as Apple AirTags to simulate mailing scenarios. By leveraging unencrypted JSON data from the AirTags' items.data files, the study captured spatial and temporal dynamics, including location accuracy and timeliness of updates, as well as metadata like battery levels and device health. These trackers provided valuable insights into logistical workflows, particularly for localized delivery scenarios, where origin and destination spokes are closely spaced but rely on a distant hub for sorting and routing.

The results highlighted critical inefficiencies in traditional hub-and-spoke systems when applied to localized delivery needs. In simulations, routing mail through a centralized hub introduced significant delays, with mail between nearby spokes taking unnecessary detours of nearly 100 miles despite their proximity. These findings suggest that centralized hubs are poorly suited for short-distance spoke-to-spoke exchanges and emphasize the need for adaptive routing strategies. The potential benefits of direct spoke-to-spoke routing or decentralized mini-hubs were demonstrated as viable solutions to reduce travel times and operational delays, especially during peak traffic periods when hub queues exacerbated inefficiencies.

The discrete event simulation (DES) model proved effective in analyzing the complex interplay of spatial configurations, demand dynamics, and processing constraints in hub-and-spoke systems. By modeling mail generation as a Poisson process and incorporating queueing theory to account for hub congestion, the DES framework revealed a saddle point in the cost function. At this optimal configuration, the balance between centralized routing efficiency and interaction demand ensured the system's scalability without excessive delays or detours.

In real-world scenarios, such as disaster relief efforts in western North Carolina during Hurricane Helene, AirTags demonstrated their utility in tracking the movement of supplies through temporary postal services. By leveraging the dense Find My network, which operates independently of traditional GPS, these trackers provided consistent and accurate location updates, even in challenging environments. Real-time data allowed for dynamic rerouting around flooded roads and infrastructure disruptions, enhancing the efficiency and effectiveness of disaster response logistics. These results suggest that BLE trackers could serve as vital tools for optimizing last-mile delivery, identifying bottlenecks, and improving resource allocation in both routine and emergency scenarios.

This study contributes to understanding the limitations of hub-dominated logistics systems and underscores the potential of BLE-based tracking technologies to address inefficiencies in last-mile delivery networks. The findings provide a foundation for future research into hybrid routing strategies and adaptive logistics systems capable of meeting the spatial and temporal challenges of dynamic delivery networks.

**ACKNOWLEDGEMENTS**

The authors thank the PeopleTec Technical Fellows program for research support.

**SUPPLEMENTAL MATERIAL 1.**

*Hardware Specifications:* Apple AirTag's BLE and NFC sleep consumption is 2.3µA giving over 10 years of "potential" battery life if not used much. The 1 year plus battery life specs is for very heavy use to cover everyone. AirTags [34] use nRF52832 SoCs built around a 32-bit Arm Cortex-M4F CPU with 512kB + 64kB RAM. There are 3 antennas used in the AirTag. Bluetooth Low Energy - 2.4GHz NFC - 13.56MHz Ultra-Wideband - 6.5-8GHz with a 500MHz Bandwidth.

| Specification | Details |
| --- | --- |
| **Model** | A2187 |
| **FCC ID** | BCGA2187 |
| **IC** | 579C-A2187 |
| **Wireless Communication** | **Bluetooth LE:** |
| | - Band: 2.4 GHz (2402-2480 MHz) |
| | - Peak Power (1 Mbps): 4.50 dBm (2.82 mW) |
| | - Peak Power (2 Mbps): 4.47 dBm (2.80 mW) |
| | **NFC:** Supported- 13.56MHz |
| | **UWB:** Supported - 6.5-8GHz with a 500MHz Bandwidth |
| **Antenna** | Integrated IFA (Inverted-F Antenna) |
| | Maximum Gain: -3.2 dBi |
| | Not user-accessible |
| **Power** | Battery: User replaceable |
| | No external/AC power required |
| **Physical Dimensions** | Diameter: 31.9 mm (1.26 inches) |
| | Thickness: 8.0 mm (0.31 inches) |
| | Weight: 11 grams (0.39 ounces) |
| **Networking/Protocol** | Compatible with Apple Find My network |
| | BLE advertising for discovery |
| | Secure element/cryptography details not provided |
| **Certifications** | FCC Part 15 Subpart C |
| | ISED RSS-247 Issue 2 |
| | ISED RSS-GEN Issue 5 |

AirTag's operational workflow begins with initial pairing through Near-Field Communication (NFC) at a frequency of 13.56 MHz, which activates the device when brought into proximity with a compatible iPhone. Following setup, the AirTag uses Bluetooth Low Energy (BLE) operating in the 2.4 GHz frequency band to broadcast an encrypted identifier continuously. This signal has an effective range of approximately 10 to 30 meters indoors and up to 100 meters in ideal conditions. Nearby Apple devices within the Find My network detect this BLE signal and anonymously relay location data to Apple's servers, enabling global tracking. During an active search, the AirTag engages its Ultra-Wideband (UWB) antenna, which operates between 6.24 and 8.24 GHz, providing precision tracking with centimeter-level accuracy over a shorter range of approximately 10 to 15 meters. This combination of NFC for initialization, BLE for wide-area tracking, and UWB for precise localization ensures efficient operation while balancing power consumption and performance.

The AirTags were paired with the iPhone through Bluetooth and assigned unique identifiers linked to a single Apple ID. As global tracking depends on relays within the Find My network, BLE 5 button-sized devices are optimized for low Size, Weight, Power, and Cost (SWaP-C) operations, enabling AirTags to achieve months of operation on a single CR2032 coin cell battery while maintaining an extremely compact form factor (32mm x 8mm). The use of BLE 5 minimizes energy expenditure by relying on periodic, short-duration transmissions rather than continuous communication. This configuration enabled passive relay tracking (depending on anonymous iPhone users passing within 100m of the AirTag). The AirTags were deployed in an initial experiment to simulate mailing scenarios. The MacBook served as the primary tool for data extraction, where the items.data file was accessed and analyzed. The JSON logs detailed the spatial and temporal dynamics of the AirTags, with fields such as location|horizontalAccuracy and location|verticalAccuracy providing a measure of spatial precision, while location|timeStamp allowed for tracking the frequency and timeliness of updates. Device metadata, including batteryLevel, deviceDiscoveryId, and serialNumber, ensured the health and functionality of the trackers were continuously monitored throughout the study.

*Software Specifications:* As shown stepwise and in pseudo-code below, the simulation environment follows SimPy's [35] implementation of a Discrete Event Simulation (DES). In a hub-and-spoke logistics network, the simulation uses Poisson sampling to model the arrival of mail at spokes, capturing demand patterns based on historical data or expected rates ($\lambda$). At the hub, queueing theory represents the delays introduced by limited processing capacity ($\mu$) and the volume of mail arrivals ($\lambda$), allowing for realistic modeling of congestion dynamics. Performance metrics such as average waiting time, queue length, and system utilization are calculated to provide insights into potential bottlenecks and inefficiencies. **For example, when the hub-to-spoke distance ($d_h$) is large and the hub operates near its capacity, significant queuing delays can arise, amplifying the inefficiency of the system.** The combination of Poisson sampling for stochastic mail arrivals and queueing models for processing dynamics enables a detailed analysis of how delays propagate through the network, offering valuable insights into the performance of hub-and-spoke systems with closely spaced origin and destination points but distant hubs.

1. **Processes:**
   - **mail_generator:** Simulates the random arrival of mail at the spokes using a Poisson process (exponential inter-arrival times).
   - **hub_processing:** Models mail processing delays at the hub based on queue length and processing time per mail item.
   - **delivery:** Handles the routing of mail, either direct spoke-to-spoke or through the hub, based on defined distances.
2. **Event Handling:**
   Events like mail arrivals, hub processing, and deliveries are executed asynchronously. The simulation environment advances time as needed for the next scheduled event.
3. **Metrics:**
   Metrics such as average transit time and queue performance are calculated after the simulation completes for analysis.
4. **Parameter Exploration:**
   By varying $d_s$ (spoke-to-spoke distance) and $d_h$ (hub-to-spoke distance), the simulation evaluates the impact of spatial configurations on system performance.

This pseudocode provides a concise overview of the DES and its logic, focusing on the components essential for modeling last-mile delivery dynamics in hub-and-spoke systems.

```
1. Initialize simulation environment:
   a. Create a simulation environment object.
   b. Define global parameters:
      - Travel speed (TRAVEL_SPEED)
      - Total simulation time (SIM_TIME)
      - Spoke-to-spoke distance (d_s)
      - Hub-to-spoke distance (d_h)
      - Mail arrival rate (ARRIVAL_RATE)

2. Define simulation processes:
   a. Mail generation process:
      i.   While True:
         ii.   Generate mail arrivals using exponential inter-arrival
   times.
         iii.  Add generated mail to the hub queue.
         iv.   Record arrival time for later analysis.
```

        v.      Wait (yield) until the next mail arrival.

  b. Hub processing process:
     i. While True:
        ii.    If hub queue is not empty:
           iii.   Process one mail item.
           iv.    Introduce a processing delay based on hub constraints.
        v.     Else:
        vi.    Wait (yield) briefly before checking again.

  c. Delivery process:
     i. While True:
        ii.    If hub queue is not empty:
           iii.   Remove mail from queue.
           iv.    Calculate delivery time based on routing:
                - Direct spoke-to-spoke: Use $d_s$ for travel time.
                - Via hub: Use $2 * d_h + d_s$ for travel time.
           v.     Wait (yield) for the calculated delivery time.
        vi.    Else:
           vii.   Wait (yield) briefly before checking again.

3. Execute simulation:
  a. Initialize mail queues for hub and spokes.
  b. Start all processes:
     i. Mail generation.
     ii. Hub processing.
     iii. Delivery.
  c. Run the simulation for the specified duration (SIM_TIME).

4. Collect and analyze results:
  a. Compute key metrics:
     i. Average transit time for all mail items.
     ii. Total mail processed.
     iii. Delivery success rate.
  b. Store results for visualization and comparison.

5. Explore multiple configurations:
  a. Vary spoke-to-spoke distance ($d_s$) and hub-to-spoke distance ($d_h$).
  b. Repeat steps 1-4 for each configuration.

**SUPPLEMENTAL MATERIAL 2. AirTags Tracking History**

| Device | ID name | Distance km | Records | Unique locations | Days active |
|---|---|---|---|---|---|
| 15 | Lara Croft | 10916.5 | 2447 | 2282 | 46 |
| 8 | Gulliver | 9121.2 | 2288 | 2154 | 43 |
| 19 | Meriwether Lewis | 7188.7 | 2006 | 1852 | 42 |
| 10 | Jack Sparrow | 5453.6 | 2569 | 2428 | 45 |
| 16 | Leif Erikson | 5447.9 | 2147 | 2010 | 43 |
| 22 | Sir Edmund Hillary | 5429.8 | 1543 | 1407 | 45 |
| 11 | Jacques Cousteau | 5128.6 | 2453 | 2336 | 45 |
| 21 | Roald Amundsen | 4609.8 | 2298 | 2129 | 46 |
| 12 | Jane Goodall | 4428.4 | 1589 | 1466 | 43 |
| 13 | King Arthur | 3029.1 | 2463 | 2332 | 46 |
| 9 | Indiana Jones | 1053.1 | 1907 | 1762 | 46 |
| 1 | Amerigo Vespucci | 816.7 | 1419 | 1333 | 45 |
| 14 | Kubla Khan | 284.4 | 1228 | 1169 | 46 |
| 0 | Amelia Earhart | 137.8 | 1335 | 1205 | 44 |
| 18 | Marco Polo | 78.56 | 1031 | 957 | 44 |
| 6 | Dora the Explorer | 48.2 | 4297 | 4233 | 46 |